\documentclass{article}

\usepackage[final]{neurips_2026}
\makeatletter
\providecommand{\@trackname}{}
\makeatother

\usepackage[utf8]{inputenc}
\usepackage[T1]{fontenc}
\usepackage{hyperref}
\usepackage{url}
\usepackage{booktabs}
\usepackage{amsfonts}
\usepackage{amsmath}
\usepackage{graphicx}
\usepackage{algorithm}
\usepackage{algpseudocode}
\usepackage{microtype}
\usepackage{xcolor}
\usepackage{placeins}
\usepackage{multirow}
\usepackage{graphicx}
\usepackage{subcaption}

\newcommand{\epstheory}{\varepsilon_{\mathrm{theory}}}
\newcommand{\epslower}{\varepsilon_{\mathrm{lower}}}

\newcommand{\mueff}{\mu_{\mathrm{eff}}}

\title{Canaries in the Bank: Auditing User-Level Privacy in Private Evolution}

\author{
  Sai Aparna Aketi \qquad \quad Enayat Ullah \qquad \quad Shripad Gade \\
  Meta Platforms, Inc.\\
  Menlo Park, USA. \\
  \texttt{\{aketiaparna, enayat, shripadgade\}@meta.com}
}

\begin{document}

\maketitle

\begin{abstract}
 Private Evolution (PE) generates high-fidelity synthetic data in federated settings without exposing users’ raw data. It aggregates clipped user votes over a shared candidate bank into a differentially private histogram, with noise
  calibrated to the worst-case user contribution. However, it is unclear whether an adversary can realize this worst-case privacy loss while following the PE protocol. We introduce a protocol-aware empirical audit in which the server
  commits to a single shared candidate bank and replaces roughly $1\%$ of its entries with \textit{probes} derived from a known, non-private \textit{canary}. We evaluate eight attacks, including an unchanged-bank baseline, exact copies,
  plausible paraphrases, and high-entropy synthetic nonces. Experiments on Yelp and Sentiment140 show that natural-text attacks remain substantially below the theoretical DP bound, while nonce-based attacks yield considerably stronger bounds
  and come closest to the mechanism’s privacy ceiling. These results quantify the gap between formal worst-case privacy and leakage achievable through protocol-valid candidate-bank manipulation.
\end{abstract}

\vspace{-20pt}
\section{Introduction}
\vspace{-5pt}
Synthetic data can reduce reliance on direct access to sensitive records while retaining information useful for downstream analysis \citep{jordon2022synthetic}. However, synthetic data is not inherently private but can still reveal sensitive information from the source data, motivating mechanisms with formal differential privacy guarantees \citep{cormode2025synthetic}.

Private Evolution produces synthetic data without training a model on private examples. In each round, a server proposes public candidates, clients vote for nearby candidates using private records, and a differentially private (DP) histogram determines which candidates survive \citep{lin2024private,xie2024private,hou2024pretext}. Because the released histogram is user-level DP, its Gaussian noise is calibrated to the worst contribution of any admissible user. This raises practical questions: (1) \emph{While following the PE protocol, can an honest-but-curious server design the candidate bank so that one user's participation is as detectable as worst-case theory permits?} (2) \emph{Does realistic user data make it substantially harder to detect user participation?} 

Our goal is to build an operational audit of PE that is both informative and minimally invasive. The auditor replaces only a fixed, pre-declared fraction of the shared candidate bank with \textit{canary probes} and otherwise leaves the mechanism unchanged. This restricted intervention creates a controlled membership signal while limiting disruption to output quality. The releases yield a confidence-bounded lower bound on attack-achievable privacy and thus directly quantifying its gap from $\epstheory$.

We present a protocol-aware audit comprising eight attack variants. All variants commit to one candidate bank before user sampling, and range from an unchanged-bank baseline and exact copies to plausible paraphrases and synthetic nonce probes that calibrate the strongest attainable signal. 
The final hypothesis test accounts for hidden participation and reports a confidence-bounded empirical $\varepsilon$ on two user-partitioned text datasets.

\paragraph{Related Work.} \textit{Privacy Auditing} or \textit{Empirical DP} is concerned with constructing privacy attacks i.e.  datasets and events whose observed probabilities contradict or approach a claimed privacy curve \citep{jagielski2020auditing,nasr2023tight,tramer2022debugging}. Federated audits such as CANIFE and one-shot privacy estimation insert optimized or random canary clients while limiting interference with ordinary client updates \citep{maddock2023canife,andrew2024oneshot}. Our setting differs in that the adversarial object is not a client update or individualized model: it is one public candidate bank sent unchanged to all clients.

Audits of DP synthetic data study worst-case datasets and disclosure from generated outputs \citep{annamalai2024theory,meeus2025canary,sun2025synbench}. SnapAudit separates deterministic computation from DP noise and searches for embedding signals in private in-context learning \citep{xia2025snapaudit}. To the best of our knowledge, no prior work studies the audit mechanisms for synthetic data generated via private evolution. 

\vspace{-5pt}
\section{Methodology}
\vspace{-5pt}
Let $S=\{s_1,\ldots,s_p\}$ be a public bank of $p$ candidate samples, and let $D_u$ contain the capped records of user $u$. An embedding model maps records and candidates to a shared representation space. For each record $x$, the assignment vector $a_S(x)\in\{0,1\}^p$ identifies its top-$k$ nearest candidates.
After normalizing each user's histogram and clipping it to threshold $C$, the user's contribution is
\begin{equation}
  \bar h_u(S)=\frac{1}{|D_u|k}\sum_{x\in D_u}a_S(x),\qquad
  v_u(S)=\bar h_u(S)\min\!\left\{1,\frac{C}{\|\bar h_u(S)\|_2}\right\}.
  \label{eq:contribution}
\end{equation}
For independent participation indicators $I_u\sim\mathrm{Bernoulli}(q)$, the released histogram is $Y=\sum_u I_u v_u(S)+Z$, where $Z\sim\mathcal N(0,(\sigma C)^2I)$
and $\sigma$ is the Gaussian noise multiplier.
In round $t$, PE releases a DP histogram $Y^{(t)}$ over the current synthetic population $S^{(t)}$. PE uses this histogram feedback to generate new samples and appends them to the top-$\tilde{k}$ samples from $S^{(t)}$ to form $S^{(t+1)}$. Every histogram release consumes privacy budget whereas selection and generation from the released histogram are post-processing.
See Algorithm~\ref{alg:private-evolution} for the complete pseudo-code of multi-round PE.

\textbf{Threat Model.}
We assume an \textbf{honest-but-curious} server (or auditor) that follows the PE protocol faithfully but uses the released DP histogram to infer whether a user participated. 
We assume that the server can insert \textit{canary} users to the protocol.

Further, it may inject $r$ \textit{probes} into the candidate bank, but commits to one ordered bank before the hidden inclusion draw. Every client receives the same bank. 
The server only observes aggregate messages from users but cannot inspect private traffic, individualize queries and alter any other specifics of the algorithm.
We also assume access to an auxiliary public sample of $n_{\mathrm{aux}}$ users drawn from the same distribution as the $N$ eligible non-canary users. The axillary set represents the public knowledge the server may have and acts as proxy to the private data distribution.

\textbf{Protocol-Aware Canary and Probe Design.}
 We consider two canary types: a randomly selected real user and a synthetic user whose records are high-entropy \textit{nonce}\footnote{A nonce is a pseudo-random number used only once in a cryptographic communication or security protocol.} samples. 
Table~\ref{tab:attacks} gives the eight attack variants. Each attack is evaluated on an independently selected set of five canaries. For the nonce variants, the synthetic users' samples and candidate pools are generated independently of natural user data.

\begin{table}[t]
  \caption{The eight audit attack variants. Nonce samples are independently generated high-entropy strings unrelated to natural text. Appendix~\ref{app:attacks} describes each construction in detail.}
  \label{tab:attacks}
  \centering
  \scriptsize
  \begin{tabular}{@{}lll@{}}
    \toprule
    Attack variant & Canary type & Candidate selection \\
    \midrule
    ordinary & random real user & use the original bank; inject no probes \\
    exact & random real user & inject exact copies of the canary records \\
    paraphrase & random real user & inject paraphrases spread across canary records \\
    paraphrase-norm & random real user & select paraphrases maximizing canary probe votes \\
    paraphrase-$\mueff$ & random real user & maximize canary votes and minimize user votes on probes \\
    nonce & synthetic nonce user & inject a random subset of nonce probes \\
    nonce-norm & synthetic nonce user & select nonce probes maximizing canary probe votes \\
    nonce-$\mueff$ & synthetic nonce user & maximize canary votes and minimize user votes on probes \\
    \bottomrule
  \end{tabular}
\end{table}

\vspace{-5pt}
\subsection{Audit Mechanism}
\vspace{-5pt}
We setup the hypothesis testing framework underlying differential privacy. Under $H_0$, a known canary user is absent from the eligible user pool; under $H_1$ it is present but participates only with probability $q$.
The audit proceeds in five steps. All attack-design and calibration choices are
  fixed using public auxiliary data before the canaries are evaluated.

  \textbf{Step 1: Construct and freeze the probe bank.}
  Let $S$ and $P$ denote the shared candidate bank and its designated probe
  coordinates respectively. During attack design, a background-aware greedy procedure selects
  probe candidates -- see Table \ref{tab:attacks} for variants. After
  each candidate is added, we recompute the canary's exact top-$k$ routing (see Appendix \ref{alg:greedy-probes}). On completion, $S$ and $P$ are frozen.

  Let  $Y_P=[Y]_P$
  be the released histogram restricted to $P$. For user $u$, let $v_{u,P}=[v_u(S)]_P$
  be the user's contribution on these coordinates \emph{after} clipping, and write
  $v_{\mathrm{can},P}$ for the canary's contribution.

  \textbf{Step 2: Estimate the canary-free background.}
  Using only the
  auxiliary public samples, we estimate
  \begin{equation}
    m_0=\mathbb E[Y_P\mid H_0],\quad
    \Sigma_P=\operatorname{Cov}(Y_P\mid H_0)
    =\underbrace{\Sigma_{\mathrm{users},P}}_{\text{random user participation}}
    +\underbrace{(\sigma C)^2I}_{\text{DP noise}}.
    \label{eq:background-moments}
  \end{equation}
  Here, $\Sigma_{\mathrm{users},P}$ captures variation due to the random
  participation of eligible non-canary users. 

  \textbf{Step 3: Score a participating canary.}
  Conditional on the canary participating, its contribution shifts the mean by
  $v_{\mathrm{can},P}$. Under a shared-covariance Gaussian approximation, the
  corresponding log-likelihood ratio is
  \begin{equation}
    \ell(Y_P)
    =v_{\mathrm{can},P}^{\top}\Sigma_P^{-1}
      \left(Y_P-m_0-\frac{v_{\mathrm{can},P}}{2}\right),
    \qquad
    \mueff
    =\sqrt{
      v_{\mathrm{can},P}^{\top}
      \Sigma_P^{-1}
      v_{\mathrm{can},P}
    }.
    \label{eq:llr}
  \end{equation}
  The quantity $\mueff$ measures the canary's separation from the background and
  is the objective maximized during greedy probe construction in Table \ref{tab:attacks}. See Appendix \ref{app:math} for details on the derivation.

  \textbf{Step 4: Marginalize over hidden participation.}
  Inclusion does not guarantee participation: under $H_1$, the canary
  participates independently with probability $q$. Marginalizing over this hidden
  participation event gives the mixture log-likelihood ratio
  \begin{equation}
    \ell_{\mathrm{mix}}(Y_P)
    =\log\!\left[(1-q)+q\exp\{\ell(Y_P)\}\right].
  \end{equation}

  \textbf{Step 5: Calibrate the decision rule and lower-bound privacy loss.}
  A calibration split fixes the decision threshold before final evaluation. On
  the held-out evaluation split, one-sided Clopper--Pearson confidence bounds on
  the confusion rates yield
  \begin{equation}
    \epslower=\max\!\left\{
      0,\,
      \log\frac{\mathrm{TPR}_{L}-\delta}{\mathrm{FPR}_{U}},\,
      \log\frac{\mathrm{TNR}_{L}-\delta}{\mathrm{FNR}_{U}}
    \right\}.
    \label{eq:epsilon}
  \end{equation}

  For each predeclared attack variant, we control the family-wise error rate at
  $\alpha=0.05$ across its five canaries. Each canary receives confidence budget
  $\alpha/5=0.01$, which is then divided equally among its four one-sided
  confidence bounds. We treat attack variants as separate predeclared experiments
  and therefore do not apply an additional simultaneous correction across
  variants.

  Finally, we compute the matching theoretical guarantee $\epstheory$ using
  Opacus's \texttt{PRVAccountant} implementation of numerical PRV composition
  \citep{gopi2021numerical,yousefpour2021opacus}, with the same participation
  probability $q$, noise multiplier $\sigma$, and failure probability $\delta$.

\vspace{-8pt}
\section{Experiments, Results \& Discussions}
\vspace{-5pt}

We use Yelp Open Dataset reviews \citep{yelpOpenDataset} and Sentiment140 tweets \citep{go2009twitter} that are naturally partitioned by users. Yelp users have 64 reviews and Sentiment140 users have 32--64 tweets. All canaries have 64 records. We use \texttt{all-MiniLM-L6-v2} with 384-dimensional L2-normalized embeddings and Euclidean top-$k$ voting \citep{allminilml6v2} with $p=8{,}192$ candidates where $k=5$, and $r=64$ probes. For each experiment, Qwen3-235B-A22B-Instruct-2507
\citep{qwen3instruct2507} generates the bank only using label information. We use  dataset-splits with 500 user cohorts each for bank design and calibration and 1,000 user cohort for final testing.

Unless stated otherwise, $C=0.1$, $\sigma=1$, $q=0.1$, $\delta=10^{-5}$, with 100 participating users per round; we report the 95\%-confidence lower bound $\epslower$ and matching $\epstheory$.

\begin{table}[h]
  \centering
  \caption{$\epslower$ for attacks with the label-only bank. Each entry is the maximum over five canaries with $\alpha=0.05$, each evaluated at $\alpha/5$ with $1M$ trials per hypothesis.}
  \label{tab:results}
  \scriptsize
  \setlength{\tabcolsep}{2.2pt}
  \begin{tabular}{@{}lcccccccc|c@{}}
    \toprule
    Dataset & ordinary & exact & paraphrase & paraphrase-norm & paraphrase-$\mueff$ & nonce & nonce-norm & nonce-$\mueff$ & $\epstheory$\\
    \midrule
    Yelp Open Dataset & 0.268 & 0.371 & 0.276 & 0.316 & 0.560 & 1.049 & \textbf{1.110} & \textbf{1.110} & 1.695\\
    Sentiment140 & 0.252 & 0.363 & 0.599 & 0.327 & 0.682 & 1.052 & \textbf{1.088} & 1.086 & 1.695\\
    \bottomrule
  \end{tabular}
\end{table}

\begin{figure}[H]
  \centering
  \includegraphics[width=0.75\textwidth]{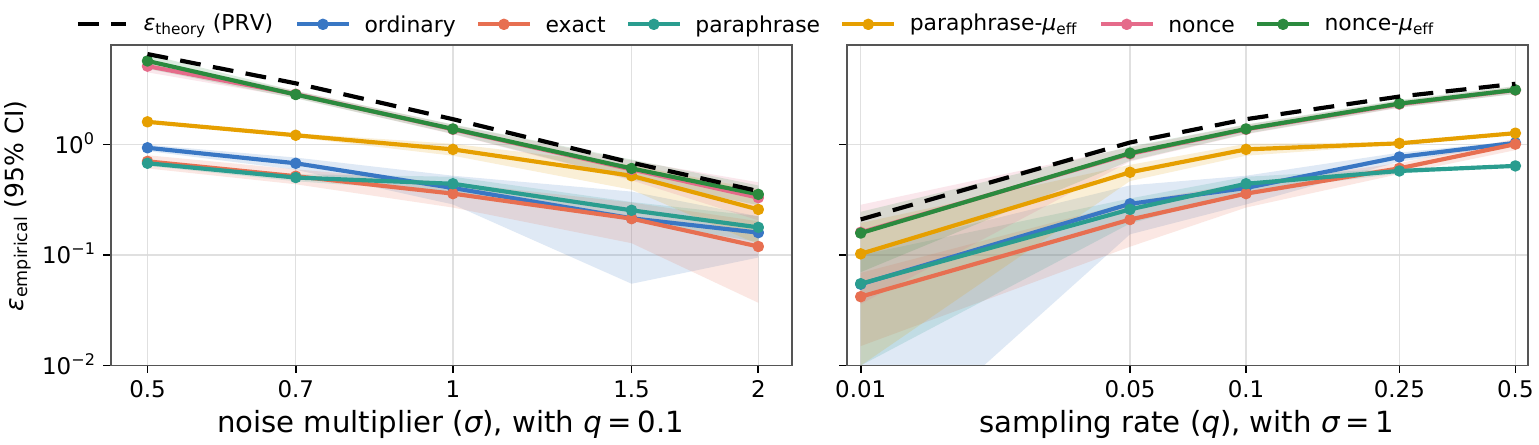}
  \caption{Sentiment140 ablations (one canary, label-only bank, $5M$ trials); bands are 95\% CIs.}
  \label{fig:ablations}
\end{figure}

Table~\ref{tab:results} reports $\epslower$ for all eight attack variants at $q=0.1$ and $\sigma=1$, against a theoretical budget of $\epstheory=1.695$. Attacks constructed from natural user text remain well below the theoretical value. We observer that the passive \texttt{ordinary} baseline attains $\epslower=0.268$ and $0.252$ on Yelp and Sentiment140, respectively ($\approx\!15\%$ of $\epstheory$), and the strongest attack with natural text, \texttt{paraphrase-}$\mueff$, attains $0.560$ and $0.682$ (below $40\%$ of $\epstheory$). Attacks based on high-entropy nonce records recover a substantially larger fraction of the budget: \texttt{nonce-norm} attains $1.110$ and $1.088$ ($\sim67\%$ of $\epstheory$). These results indicate that, although the PE noise is calibrated to the worst-case user contribution, protocol-valid manipulation of the candidate bank with natural text realizes only a small fraction of the permitted privacy loss, whereas synthetic nonce probes approach the theoretical limit. The same trends holds across noise multipliers and sampling rates (Figure~\ref{fig:ablations}).

\begin{figure}[h]
  \centering
  \includegraphics[width=0.68\textwidth]{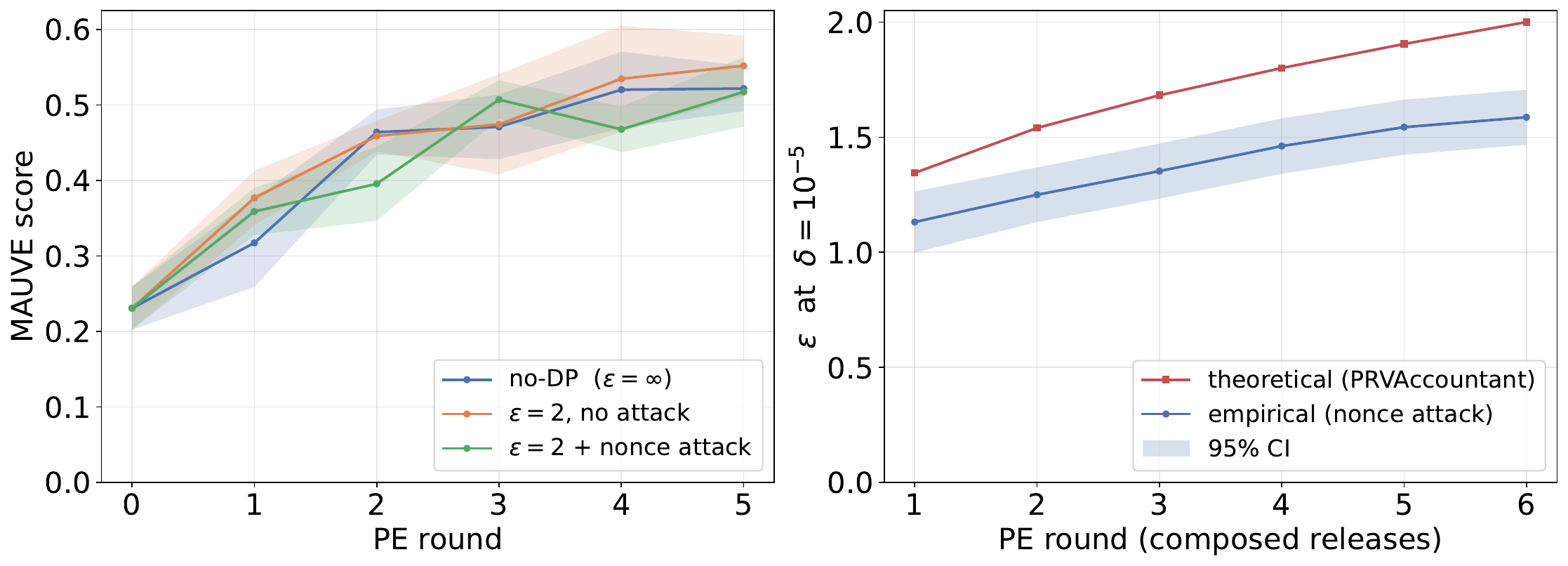}
  \caption{Multi-round PE on Yelp: utility and privacy parameters (theory vs audit) against PE-round.}
  \label{fig:multiround}
\end{figure}

\textbf{Multi-round PE with ICL:} Figure~\ref{fig:multiround} reports MAUVE utility \citep{pillutla2021mauve} and privacy across PE rounds. Each round retains the top 5\% of candidates and the probes, generates the next bank with ICL, and repeats the audit. We show the audit preserves synthetic-data utility across PE rounds while providing reasonable empirical epsilon lower bounds.

\textbf{Embedding comparison:} Table~\ref{tab:embedding-models} varies the encoder across three sentence-transformer models of increasing capacity: MiniLM-L6-v2 ($22.7$M parameters, $384$-d), MiniLM-L12-v2 ($33.4$M, $384$-d), and MPNet-base-v2 ($109$M, $768$-d). We observe that the empirical epsilon of PE is insensitive to encoder capacity for this setup. 
\begin{table}[!h]
    \centering
    \caption{Embedding model ablation ($\epslower$).}
    \label{tab:embedding-models}
    \scriptsize
    \setlength{\tabcolsep}{2.2pt}
    \begin{tabular}{@{}lcccccc@{}}
    \toprule
    \multirow{2}{*}{Attack} & \multicolumn{3}{c}{Yelp Open Reviews} & \multicolumn{3}{c}{Sentiment140 Tweets} \\
    \cmidrule(lr){2-4} \cmidrule(lr){5-7}
    & MiniLM-L6 & MiniLM-L12 & MPNET-Base-V2 & MiniLM-L6 & MiniLM-L12 & MPNET-Base-V2 \\
    \midrule
    ordinary & 0.268 & 0.232 & 0.273 & 0.252 & 0.272 & 0.249 \\
    exact    & 0.371 & 0.357 & 0.387 & 0.363 & 0.341 & 0.245 \\
    nonce    & 1.049 & 1.105 & 1.087 & 1.052 & 1.088 & 1.078 \\
    \bottomrule
\end{tabular}
\end{table}

\vspace{-12pt}
\section{Conclusion and Limitations}
\vspace{-8pt}
In this work, we designed a privacy auditing procedure for private evolution. Experiments show that practical attacks fall well short of worst-case theory, but (high entropy) nonce datasets bring empirical epsilon closer to theoretical epsilon. We note that designing and running a tight audit requires multiple trials, making it computationally expensive to run in production settings. In the future, we will explore design of \textit{one run auditing} \citep{steinke2023privacy} procedures to trade-off runtime and privacy measurement.

\FloatBarrier
\bibliographystyle{plainnat}
\bibliography{references}

\clearpage
\appendix
\numberwithin{algorithm}{section}

\section{Multi-Round Private Evolution Algorithm}
\label{app:private-evolution}

Algorithm~\ref{alg:private-evolution} presents multi-round PE independently of our experimental parameter choices. Each round computes a user-level DP histogram over the current synthetic population. If $J^{(t)}=\operatorname{Top}_{\tilde{k}}(Y^{(t)})$, PE retains $R^{(t)}=\{s_j^{(t)}:j\in J^{(t)}\}$ and constructs generation feedback $F^{(t)}$ from $S^{(t)}$ and $Y^{(t)}$. A public generator produces $V^{(t)}=\mathcal G(F^{(t)})$, and PE sets $S^{(t+1)}=R^{(t)}\cup V^{(t)}$. The feedback rule is method-dependent and may request paraphrases or supply positively and negatively scored ICL examples. The private data enter only through the bounded client contributions in Lines~\ref{line:client-start}--\ref{line:client-end}. Subsequent selection and generation use only the DP histogram and public information, so they are post-processing. Because the candidate population evolves between rounds, the accountant must compose all $T$ histogram releases under the declared sampling rule.

\begin{algorithm}[h]
\caption{Multi-round Private Evolution with user-level DP top-$k$ feedback}
\label{alg:private-evolution}
\begin{algorithmic}[1]
\Require Private user datasets $\{D_u\}_{u\in\mathcal U}$; initial public population $S^{(0)}$
\Require encoder $\phi$ and distance $d$; top-$k$ rule; user cap $m$
\Require clipping threshold $C$; noise multiplier $\sigma$; user sampler $\mathsf{SampleUsers}$
\Require rounds $T$; retention count $\tilde{k}$; number of variations $g$; public ICL generator $\mathcal G$
\For{$t=0,\ldots,T-1$}
  \State Let $p_t\gets|S^{(t)}|$ and publish the same ordered bank $S^{(t)}$ and encoder $\phi$ to all sampled users
  \State $B^{(t)}\gets\mathsf{SampleUsers}(\mathcal U)$
  \ForAll{$u\in B^{(t)}$ \textbf{in parallel}} \label{line:client-start}
    \State $D'_u\gets\mathsf{CapRecords}(D_u,m)$ and $h_u\gets\mathbf 0_{p_t}$
    \ForAll{$x\in D'_u$}
      \State $N_k(x)\gets\underset{J\subseteq[p_t],\,|J|=k}{\arg\min}\ \sum_{j\in J}d(\phi(x),\phi(s^{(t)}_j))$
      \ForAll{$j\in N_k(x)$}
        \State $h_{u,j}\gets h_{u,j}+1$
      \EndFor
    \EndFor
    \State $\bar h_u\gets h_u/(|D'_u|k)$ \Comment{by-vote normalization}
    \State $v_u\gets\bar h_u\min\{1,C/\|\bar h_u\|_2\}$ \Comment{user-level clipping}
    \State Send $v_u$ to the protected aggregator \label{line:client-end}
  \EndFor
  \State $Y^{(t)}\gets\sum_{u\in B^{(t)}}v_u+Z^{(t)}$, where $Z^{(t)}\sim\mathcal N(0,(\sigma C)^2I_{p_t})$
  \State $J^{(t)}\gets\operatorname{Top}_{\tilde{k}}(Y^{(t)})$ \Comment{indices selected by the DP histogram}
  \State $R^{(t)}\gets\{s^{(t)}_j:j\in J^{(t)}\}$ \Comment{retain top previous samples}
  \State $F^{(t)}\gets\mathsf{BuildFeedback}(S^{(t)},Y^{(t)})$ \Comment{e.g., paraphrases or positive/negative ICL}
  \State $V^{(t)}\gets\mathcal G(F^{(t)};g)$ \Comment{generate $g$ new samples}
  \State $S^{(t+1)}\gets R^{(t)}\cup V^{(t)}$ \Comment{retain and append}
\EndFor
\State \Return $S^{(T)}$ \Comment{evolved synthetic population}
\end{algorithmic}
\end{algorithm}

\section{Attack Variant Details}
\label{app:attacks}

All variants preserve one ordered bank shared by every participating user. Active variants reserve an ordered list of $r=64$ replacement positions before selection. Let $S_{\mathrm{base}}$ be the original bank after removing those positions. If probe $j_t$ is selected in round $t$, it is assigned to replacement position $t$; the completed bank is then frozen before membership trials. Canary, auxiliary, and final-test users are disjoint: canary records define the desired signal, auxiliary users estimate background traffic for probe selection, and final-test users only evaluate the frozen attack.

\paragraph{Candidate pools.}
Each paraphrase or nonce attack begins with an indexed pool $\mathcal Q=\{c_1,\ldots,c_M\}$ of $M=512$ candidate probes and injects $r=64$. The paraphrase pool contains non-verbatim rewrites generated across the 64 real-canary records. The nonce pool contains fresh high-entropy strings generated independently of natural text and of the synthetic nonce canary. The unoptimized variants choose 64 probes directly; the norm and $\mueff$ variants use the greedy procedure in Algorithm~\ref{alg:greedy-probes}.

\paragraph{Trial-set routing and scores.}
For a trial index set $P$, define $S(P)=S_{\mathrm{base}}\cup\{c_j:j\in P\}$. For each of the 64 canary records, we recompute its five nearest entries in $S(P)$. It is sufficient to merge the trial probes with that record's five nearest entries in $S_{\mathrm{base}}$: no lower-ranked base entry can enter the final top five. Let $w_P$ be the resulting vector of probe-vote counts divided by the canary's $64\times5=320$ total votes. The two optimized attacks differ only in the scalar score
\begin{equation}
  J_{\mathrm{norm}}(P)=\|w_P\|_2^2,
  \qquad
  J_{\mueff}(P)=w_P^\top\widehat\Sigma_P^{-1}w_P,
  \label{eq:greedy-probe-scores}
\end{equation}
Before selection, we estimate a covariance $\widehat\Sigma_{\mathcal Q}$ over the full 512-candidate pool from the fixed auxiliary-user split and freeze it. A trial set $P$ uses the corresponding principal submatrix $\widehat\Sigma_P=[\widehat\Sigma_{\mathcal Q}]_{P,P}$. Thus the matrix used by the score changes with the selected coordinates, but it is not re-estimated from auxiliary users at every step. Norm rewards concentrated canary votes, while $\mueff$ discounts probe coordinates that also receive background-user votes. The mechanism's full-vector clipping is still applied when the frozen bank is audited. Appendix~\ref{app:hyperparameters} gives the estimator and shrinkage coefficient.

In Algorithm~\ref{alg:greedy-probes}, $\mathsf{ProbeVotes}(S(P),D_{\mathrm{can}},k)$ denotes the full re-routing operation above and returns the normalized vector $w_P$. It recomputes all 64 canary records rather than updating the previous round's counts.

\paragraph{Why forward selection is necessary.}
Probes compete for the same top-5 votes. Adding a closer probe can take a vote from an already selected probe, so the value of a probe depends on the rest of the selected set. Ranking all 512 candidates independently and taking the top 64 is therefore incorrect. At round $t$, the algorithm adds each remaining candidate to the $t-1$ selected probes, recomputes all canary routes and the complete score, and commits the best trial set. Exact score ties are resolved by the lowest pool index for both objectives. This deterministic rule prevents arbitrary ties from creating a false difference between nonce-norm and nonce-$\mueff$ when their objectives are equivalent.

\begin{algorithm}[H]
\caption{Greedy construction of norm and $\mueff$ probe sets}
\label{alg:greedy-probes}
\begin{algorithmic}[1]
\Require Non-probe bank $S_{\mathrm{base}}$; indexed pool $\mathcal Q=\{c_j\}_{j=1}^{M}$ with $M=512$
\Require Canary records $D_{\mathrm{can}}$ with $m=64$; auxiliary users $D_{\mathrm{aux}}$; $k=5$; budget $r=64$
\Require Objective $o\in\{\mathrm{norm},\mueff\}$; ordered replacement positions $(\rho_1,\ldots,\rho_r)$
\State $P\gets\emptyset$
\If{$o=\mueff$}
  \State $\widehat\Sigma_{\mathcal Q}\gets\mathsf{AuxCovariance}(\mathcal Q,D_{\mathrm{aux}})$ \Comment{estimate once and freeze}
\EndIf
\For{$t=1,\ldots,r$}
  \ForAll{$j\in[M]\setminus P$}
    \State $T\gets P\cup\{j\}$
    \State $w_T\gets\mathsf{ProbeVotes}(S(T),D_{\mathrm{can}},k)$ \Comment{recompute all top-$k$ routes}
    \If{$o=\mathrm{norm}$}
      \State $s_j\gets\|w_T\|_2^2$
    \Else
      \State $\widehat\Sigma_T\gets[\widehat\Sigma_{\mathcal Q}]_{T,T}$
      \State $s_j\gets w_T^\top\widehat\Sigma_T^{-1}w_T$
    \EndIf
  \EndFor
  \State $j_t\gets\min\!\left(\arg\max_{j\in[M]\setminus P}s_j\right)$ \Comment{lowest-index tie break}
  \State $P\gets P\cup\{j_t\}$
\EndFor
\State Insert $c_{j_t}$ at position $\rho_t$ for $t=1,\ldots,r$ and freeze the ordered bank
\State \Return frozen bank and selected sequence $(j_1,\ldots,j_r)$
\end{algorithmic}
\end{algorithm}

\paragraph{Ordinary.}
This passive baseline randomly selects a real canary user and leaves the candidate bank unchanged. The audit observes the histogram coordinates on which the canary naturally votes under the original top-$k$ routing. It measures how distinguishable the canary is without adding probes or otherwise adapting the bank to that user.

\paragraph{Exact.}
This variant uses the same randomly selected real user and injects one exact copy of each of its 64 records. Each copied record is at zero embedding distance from its source.

\paragraph{Paraphrase.}
From the 512-item paraphrase pool, this unoptimized semantic baseline selects 64 probes spread across the canary's records rather than concentrating on a few records. It uses no norm or background-aware score.

\paragraph{Paraphrase-norm.}
This variant applies Algorithm~\ref{alg:greedy-probes} to the 512-paraphrase pool using $J_{\mathrm{norm}}$. It favors sets on which canary votes have high magnitude and are concentrated on relatively few probes, without accounting for background traffic.

\paragraph{Paraphrase-$\mueff$.}
This primary realistic attack runs the same algorithm and pool with $J_{\mueff}$. Its frozen pool covariance comes only from auxiliary users, and each trial uses the submatrix for its selected coordinates; no final-test user influences selection. It favors paraphrases that attract the canary but lie on coordinates with little or uncorrelated background traffic.

\paragraph{Nonce.}
This calibration constructs a synthetic canary from 64 fresh, high-entropy nonce records and independently generates the 512-item nonce probe pool. It injects 64 pool entries selected uniformly without replacement. These are independent audit markers, not hashes of private records; they place signal where natural users are unlikely to vote and approximate the mechanism ceiling.

\paragraph{Nonce-norm.}
This variant applies Algorithm~\ref{alg:greedy-probes} to the 512-nonce pool using $J_{\mathrm{norm}}$. It estimates the mechanism ceiling when selection explicitly maximizes the magnitude and concentration of nonce-canary votes without background covariance.

\paragraph{Nonce-$\mueff$.}
This variant runs the same nonce search with $J_{\mueff}$. If nonce strings separate cleanly from natural text, $\widehat\Sigma_P$ is approximately isotropic and the selected sequence should match nonce-norm, including under ties. If natural users collide with nonce probes, the objective preserves canary votes while avoiding those background directions.

\section{Complete Results}
\label{app:full-results}

Bank A is generated by Qwen3-235B-A22B-Instruct-2507 using label-only prompts. Bank B uses the same generator with three public in-context examples, yielding candidates that are better matched to the private-user distribution. Table~\ref{tab:full-results} reports both candidate-bank arms for each dataset.

\begin{table}[h]
  \caption{$\epslower$ for two datasets and two candidate-bank arms. Each entry is the maximum over five canaries with $\alpha=0.05$, each evaluated at $\alpha/5=0.01$ and $1M$ trials per hypothesis; $\epstheory=1.695$.}
  \label{tab:full-results}
  \centering
  \small
  \begin{tabular}{@{}lrrrr@{}}
    \toprule
    Attack variant & Yelp A & Yelp B & Sentiment140 A & Sentiment140 B \\
    \midrule
    ordinary & 0.247 & 0.295 & 0.475 & 0.429 \\
    exact & 0.313 & 0.304 & 0.363 & 0.587 \\
    paraphrase & 0.276 & 0.297 & 0.599 & 0.759 \\
    paraphrase-norm & 0.316 & 0.457 & 0.327 & 0.871 \\
    paraphrase-$\mueff$ & 0.560 & 0.439 & 0.682 & 0.932 \\
    nonce & 1.038 & 1.047 & 1.018 & 1.014 \\
    nonce-norm & \textbf{1.110} & \textbf{1.120} & \textbf{1.088} & \textbf{1.063} \\
    nonce-$\mueff$ & \textbf{1.110} & 0.958 & 1.086 & 0.977 \\
    \bottomrule
  \end{tabular}
\end{table}

\section{Hyperparameter settings}
\label{app:hyperparameters}

\paragraph{Users and record cap.}
All experiments cap each user at $m=64$ records. Yelp admits users with exactly 64 records, whereas Sentiment140 admits users with 32--64 records. The single-release experiments use an eligible non-canary population of $N=1{,}000$ users and $q=0.1$ unless swept, giving 100 expected participants per release. We use 500 disjoint auxiliary users to design probes and estimate non-canary moments, 500 calibration users to lock the threshold, and the $N=1{,}000$ final-test users only for evaluation.

\paragraph{Trial budgets.}
Table~\ref{tab:results} and Table~\ref{tab:full-results} use $1M$ trials per hypothesis. Figure~\ref{fig:ablations} uses $5M$ trials per hypothesis because small noise multipliers separate the hypotheses more strongly, making errors at the selected tail threshold rare; the larger budget stabilizes the Clopper--Pearson bounds. The multi-round audit also uses $5M$ trials per hypothesis.

\paragraph{Multi-round accounting.}
Figure~\ref{fig:multiround} composes $T=6$ Poisson-sampled releases on Yelp with $q=0.1$, $\sigma=1.1088$, $\delta=10^{-5}$, $N=8{,}000$ eligible test users, and 500 auxiliary users. For each prefix $t$, a fresh Opacus \texttt{PRVAccountant} is stepped once per release with the same sampling rate and noise multiplier, then queried at $\delta=10^{-5}$. This gives $\epstheory=1.345,1.540,1.682,1.800,1.905,$ and $2.000$ after releases one through six, respectively. The empirical audit composes the same prefixes by summing their mixture log-likelihood ratios.

\section{Additional Mathematical Details}
\label{app:math}

This appendix supplies derivations omitted from the main text.  In particular,
it distinguishes the absent-canary, fixed-inclusion, and hidden-participation
distributions; derives the Gaussian and mixture likelihood ratios; records the
distribution of the resulting score; and connects the final hypothesis test to
a confidence-bounded empirical privacy lower bound.  These derivations do not
change the audit mechanism or its assumptions.

\subsection{Background aggregate and its moments}

Let \(P\) denote the histogram coordinates inspected by the audit.  Write the
contribution of non-canary user \(u\), restricted to these coordinates, as
\[
c_u=v_{u,P}.
\]
For a fixed eligible population of \(N\) non-canary users, define the
background aggregate
\begin{equation}
B=\sum_{u=1}^{N} I_uc_u+Z_P,
\qquad
I_u\overset{\mathrm{ind}}{\sim}\operatorname{Bernoulli}(q),
\qquad
Z_P\sim\mathcal N\!\left(0,(\sigma C)^2I\right).
\label{eq:section-background}
\end{equation}
The user vectors \(c_u\) are fixed after the candidate bank, routing rule, and
clipping operation have been frozen.  Randomness comes from user sampling and
Gaussian DP noise.

Since \(\mathbb E[I_u]=q\), the absent-canary mean is
\begin{equation}
m_0:=\mathbb E[B]
=q\sum_{u=1}^{N}c_u.
\label{eq:section-background-mean}
\end{equation}
For a fixed vector \(c_u\),
\[
\operatorname{Cov}(I_uc_u)
=\operatorname{Var}(I_u)c_uc_u^\top
=q(1-q)c_uc_u^\top.
\]
Independence therefore gives
\begin{equation}
\Sigma:=\operatorname{Cov}(B)
=q(1-q)\sum_{u=1}^{N}c_uc_u^\top
+(\sigma C)^2I.
\label{eq:section-background-covariance}
\end{equation}
The off-diagonal entries capture correlated voting: one user may contribute to
several inspected coordinates simultaneously.

\subsection{Three cases}

Let
\[
v=v_{\mathrm{can},P}
\]
be the known canary contribution on the inspected coordinates.  Because the
bank and canary records are fixed before the trial, \(v\) is a deterministic
vector.

\begin{center}
\begin{tabular}{llll}
\toprule
Case & Released vector & Mean & Covariance \\
\midrule
Canary absent & \(B\) & \(m_0\) & \(\Sigma\) \\
Canary sampled & \(B+v\) & \(m_0+v\) & \(\Sigma\) \\
Canary eligible & \(B+Jv\) & \(m_0+qv\) &
\(\Sigma+q(1-q)vv^\top\) \\
\bottomrule
\end{tabular}
\end{center}
Here \(J\sim\operatorname{Bernoulli}(q)\) is the canary's sampling indicator,
independent of \(B\).

The covariance is the same under absence and \emph{fixed inclusion} because
adding a deterministic vector changes only the mean:
\begin{align}
\operatorname{Cov}(B+v)
&=\mathbb E\!\left[
\bigl(B+v-(m_0+v)\bigr)
\bigl(B+v-(m_0+v)\bigr)^\top
\right] \notag\\
&=\mathbb E\!\left[
(B-m_0)(B-m_0)^\top
\right] \notag\\
&=\Sigma.
\label{eq:section-shift-covariance}
\end{align}
In contrast, the full eligible-canary hypothesis contains the random term
\(Jv\):
\begin{align}
\operatorname{Cov}(B+Jv)
&=\operatorname{Cov}(B)+\operatorname{Cov}(Jv) \notag\\
&=\Sigma+q(1-q)vv^\top.
\label{eq:section-mixture-covariance}
\end{align}
Thus the full alternative does not have the same covariance as the null.  It
is a mixture of the absent and fixed-inclusion distributions.

Independent Bernoulli sampling is important for this argument.  Adding an
eligible canary does not displace an ordinary user, so the background
distribution remains unchanged.  Under fixed-size sampling without
replacement, conditioning on canary inclusion could change the distribution
of the remaining sampled users.

\subsection{Gaussian approximation and fixed-inclusion likelihood ratio}

The background is a Bernoulli-weighted sum of user vectors convolved with
Gaussian noise.  The audit approximates it by
\[
B\approx\mathcal N(m_0,\Sigma).
\]
Consequently, the absent and fixed-inclusion component distributions are
\begin{align}
H_0 &: Y\approx\mathcal N(m_0,\Sigma), \label{eq:section-h0}\\
H_{\mathrm{inc}} &: Y\approx\mathcal N(m_0+v,\Sigma).
\label{eq:section-hinc}
\end{align}
The equality of the component covariances is exact; their Gaussian shape is
the approximation.

The fixed-inclusion log-likelihood ratio is
\[
\ell(y)
:=\log\frac{p_{\mathrm{inc}}(y)}{p_0(y)}.
\]
Using the two Gaussian densities,
\begin{align*}
\ell(y)
&=-\frac12\left[
(y-m_0-v)^\top\Sigma^{-1}(y-m_0-v)
-(y-m_0)^\top\Sigma^{-1}(y-m_0)
\right].
\end{align*}
Expanding the first quadratic form gives
\begin{align*}
(y-m_0-v)^\top\Sigma^{-1}(y-m_0-v)
&=(y-m_0)^\top\Sigma^{-1}(y-m_0)\\
&\quad-2v^\top\Sigma^{-1}(y-m_0)
+v^\top\Sigma^{-1}v.
\end{align*}
The shared quadratic term cancels, leaving
\begin{equation}
\boxed{
\ell(y)
=v^\top\Sigma^{-1}(y-m_0)
-\frac12v^\top\Sigma^{-1}v
=v^\top\Sigma^{-1}
\left(y-m_0-\frac{v}{2}\right).
}
\label{eq:section-fixed-llr}
\end{equation}
The score is linear in \(y\) precisely because the two Gaussian components
have equal covariance.

\subsection{Whitening and effective separation}

Define
\[
x=\Sigma^{-1/2}(y-m_0),
\qquad
s=\Sigma^{-1/2}v.
\]
In whitened coordinates,
\[
H_0:x\sim\mathcal N(0,I),
\qquad
H_{\mathrm{inc}}:x\sim\mathcal N(s,I),
\]
and
\[
\ell(x)=s^\top x-\frac12\lVert s\rVert_2^2.
\]
The distance between the component means is
\begin{equation}
\boxed{
\mu_{\mathrm{eff}}
=\lVert s\rVert_2
=\sqrt{v^\top\Sigma^{-1}v}.
}
\label{eq:section-mueff}
\end{equation}
This is the Mahalanobis signal-to-noise ratio.  Directions with large
background variance receive less weight, while signal on quiet coordinates
receives more weight.

Writing \(d=\mu_{\mathrm{eff}}\), the score distributions are
\begin{align}
\ell(Y)\mid H_0
&\sim\mathcal N\!\left(-\frac{d^2}{2},d^2\right),
\label{eq:section-llr-h0}\\
\ell(Y)\mid H_{\mathrm{inc}}
&\sim\mathcal N\!\left(+\frac{d^2}{2},d^2\right).
\label{eq:section-llr-hinc}
\end{align}
At threshold \(a\),
\begin{align}
\operatorname{FPR}(a)
&=1-\Phi\!\left(\frac{a+d^2/2}{d}\right),\\
\operatorname{TPR}_{\mathrm{inc}}(a)
&=1-\Phi\!\left(\frac{a-d^2/2}{d}\right).
\end{align}
At the equal-prior threshold \(a=0\),
\[
\operatorname{FPR}=\Phi(-d/2),
\qquad
\operatorname{TPR}_{\mathrm{inc}}=\Phi(d/2).
\]

Because
\[
\Sigma\succeq(\sigma C)^2I
\qquad\text{and}\qquad
\lVert v\rVert_2\le C,
\]
the conditional Gaussian separation satisfies
\begin{equation}
\mu_{\mathrm{eff}}^2
=v^\top\Sigma^{-1}v
\le\frac{\lVert v\rVert_2^2}{(\sigma C)^2}
\le\frac{1}{\sigma^2}.
\label{eq:section-mueff-ceiling}
\end{equation}
This is a ceiling on component separation, not directly an
\((\varepsilon,\delta)\)-DP formula.

\subsection{Hidden participation and the mixture likelihood ratio}

Under the actual alternative, the canary is eligible but is sampled only with
probability \(q\).  Therefore
\begin{equation}
p_1(y)=(1-q)p_0(y)+q\,p_{\mathrm{inc}}(y).
\label{eq:section-mixture-density}
\end{equation}
Dividing by \(p_0(y)\) and using
\(p_{\mathrm{inc}}(y)/p_0(y)=e^{\ell(y)}\) gives
\begin{equation}
\boxed{
\ell_{\mathrm{mix}}(y)
:=\log\frac{p_1(y)}{p_0(y)}
=\log\left[(1-q)+qe^{\ell(y)}\right].
}
\label{eq:section-mixture-llr}
\end{equation}
This expression explicitly preserves the mixture; it does not approximate
full \(H_1\) as a single Gaussian.  The identity is exact if \(\ell\) is the
true component likelihood ratio.  In the audit, it inherits the Gaussian
approximation used to obtain Equation~\eqref{eq:section-fixed-llr}.

For \(q>0\),
\[
\frac{d\ell_{\mathrm{mix}}}{d\ell}
=\frac{qe^\ell}{(1-q)+qe^\ell}>0.
\]
Hence \(\ell\) and \(\ell_{\mathrm{mix}}\) rank observations identically for a
single release, although only \(\ell_{\mathrm{mix}}\) is the likelihood-ratio
value for the eligible-canary hypothesis.

For any fixed threshold event,
\begin{equation}
\operatorname{TPR}
=(1-q)\operatorname{FPR}
+q\,\operatorname{TPR}_{\mathrm{inc}},
\label{eq:section-mixture-tpr}
\end{equation}
and therefore
\[
\operatorname{TPR}-\operatorname{FPR}
=q\left(
\operatorname{TPR}_{\mathrm{inc}}-\operatorname{FPR}
\right).
\]
This equation makes the effect of user subsampling explicit: only a fraction
\(q\) of eligible-canary trials contain the canary's signal.

\subsection{From test errors to an empirical privacy lower bound}

Let \(A\) be the event that the frozen audit predicts that the canary is
present.  Define
\[
\operatorname{TPR}=\mathbb P_{H_1}(A),
\qquad
\operatorname{FPR}=\mathbb P_{H_0}(A),
\]
and, for the complementary event,
\[
\operatorname{TNR}=\mathbb P_{H_0}(A^c),
\qquad
\operatorname{FNR}=\mathbb P_{H_1}(A^c).
\]
For neighboring user-level datasets, \((\varepsilon,\delta)\)-DP requires
\begin{align}
\mathbb P_{H_1}(A)
&\le e^\varepsilon\mathbb P_{H_0}(A)+\delta,
\label{eq:section-dp-forward}\\
\mathbb P_{H_0}(A^c)
&\le e^\varepsilon\mathbb P_{H_1}(A^c)+\delta.
\label{eq:section-dp-reverse}
\end{align}
Substitution and rearrangement yield
\begin{align}
\varepsilon
&\ge
\log\frac{\operatorname{TPR}-\delta}{\operatorname{FPR}},
\label{eq:section-eps-forward}\\
\varepsilon
&\ge
\log\frac{\operatorname{TNR}-\delta}{\operatorname{FNR}}.
\label{eq:section-eps-reverse}
\end{align}

The rates are estimated from finitely many final-evaluation trials.  Let
\(\operatorname{TPR}_L\) and \(\operatorname{TNR}_L\) denote one-sided lower
confidence bounds, and let \(\operatorname{FPR}_U\) and
\(\operatorname{FNR}_U\) denote one-sided upper confidence bounds.  A
conservative empirical privacy lower bound is
\begin{equation}
\boxed{
\varepsilon_{\mathrm{lower}}
=\max\left\{
0,\,
\log\frac{\operatorname{TPR}_L-\delta}{\operatorname{FPR}_U},\,
\log\frac{\operatorname{TNR}_L-\delta}{\operatorname{FNR}_U}
\right\}.
}
\label{eq:section-epsilon-lower}
\end{equation}
A branch whose lower-confidence numerator is at most \(\delta\) supplies no
positive bound.

For \(x\) successes among \(n\) Bernoulli trials, the one-sided
Clopper--Pearson endpoints at tail probability \(\gamma\) are
\begin{align}
p_L
&=
\begin{cases}
0, & x=0,\\
F_{\mathrm{Beta}(x,n-x+1)}^{-1}(\gamma), & x>0,
\end{cases}\\
p_U
&=
\begin{cases}
1, & x=n,\\
F_{\mathrm{Beta}(x+1,n-x)}^{-1}(1-\gamma), & x<n.
\end{cases}
\end{align}
With overall \(\alpha=0.05\), five canaries per predeclared attack, and four
one-sided rate bounds per canary, the audit uses
\[
\gamma=\frac{0.05}{5\cdot4}=0.0025.
\]
The calibration split is independent of final evaluation so that the event
\(A\), including its threshold, is frozen before the final confusion counts
are observed.

\subsection{Auxiliary moment estimation}

For \(n_{\mathrm{aux}}\) auxiliary users with clipped contributions \(c_u\) on
the inspected coordinates, define
\begin{align}
\widehat\mu_P
&=\frac{1}{n_{\mathrm{aux}}}
\sum_{u=1}^{n_{\mathrm{aux}}}c_u,\\
\widehat M_P
&=\frac{1}{n_{\mathrm{aux}}}
\sum_{u=1}^{n_{\mathrm{aux}}}c_uc_u^\top.
\end{align}
The second quantity is an uncentered second moment, rather than a sample
covariance, because Bernoulli participation contributes
\[
\operatorname{Cov}(I_uc_u)=q(1-q)c_uc_u^\top.
\]

Because $n_{\mathrm{aux}}$ is finite, off-diagonal entries of $\widehat M_P$ can be noisy. We therefore use diagonal shrinkage
\begin{equation}
  \operatorname{shrink}_{0.1}(\widehat M_P)
  =0.9\widehat M_P+0.1\operatorname{diag}(\operatorname{diag}(\widehat M_P)),
\end{equation}
which preserves each diagonal second moment and multiplies the estimated off-diagonal terms by $0.9$. 
The plug-in estimators thus used by the audit are
\begin{align}
\widehat m_0
&=qN\widehat\mu_P,\\
\widehat\Sigma_P
&=q(1-q)N\operatorname{shrink}_{0.1}(\widehat M_P)
+(\sigma C)^2I+10^{-9}I,
\end{align}

The term $(\sigma C)^2I$ is the covariance of the Gaussian DP noise already present in the released histogram; no noise is added to the public auxiliary data. The final $10^{-9}I$ term is only a numerical ridge for stable inversion.

\subsection{Conditional composition across rounds}

When the candidate bank evolves across rounds, the joint transcript likelihood
ratio follows the chain rule
\begin{equation}
\log\frac{p_1(y^{1:T})}{p_0(y^{1:T})}
=\sum_{t=1}^{T}
\log
\frac{
p_1(y^{(t)}\mid y^{(<t)})
}{
p_0(y^{(t)}\mid y^{(<t)})
}.
\label{eq:section-conditional-composition}
\end{equation}
Thus per-round mixture scores may be summed when each score is the appropriate
conditional likelihood ratio given the realized earlier transcript.  Summing
unconditional marginal likelihood ratios would not generally be correct for
an adaptive candidate-bank process.  Formal DP accounting remains valid under
adaptive composition because candidate retention and public generation are
post-processing, while every new private histogram is separately composed.

\end{document}